\RequirePackage{fix-cm}
\documentclass[twocolumn,epjc3]{svjour3}  
\smartqed  
\usepackage{cite}
\usepackage{float}
\usepackage{amsmath}
\RequirePackage{graphicx}
\usepackage{hyperref}
\hypersetup{
   colorlinks=true,
    linkcolor=blue,
    filecolor=magenta,      
    urlcolor=blue,
    pdftitle={Sharelatex Example},
    bookmarks=true,
    citecolor=blue,
    pdfpagemode=FullScreen,
}

\journalname{Eur. Phys. J. A}
\begin{document}

\title{Mass-Spectroscopy of hidden charm and hidden strange tetraquarks in diquark-antidiquark approach
}


\author{Rohit Tiwari\thanksref{e1,addr1}
        \and
         Ajay Kumar Rai\thanksref{addr1}
}

\thankstext{e1}{e-mail: rohittiwari843@gmail.com}


\institute{Department of Physics, Sardar Vallabhbhai National Institute of Technology, Surat-395007, Gujarat, INDIA. \label{addr1}}

\date{Received: date / Accepted: date}

\maketitle
\begin{abstract}
We investigated the four-quark systems with quark structures of $cs\bar{c}\bar{s}$, $cq\bar{c}\bar{s}$, $bs\bar{b}\bar{s}$, and $bq\bar{b}\bar{s}$ in the framework of the non-relativistic quark model, motivated by the recent observation of exotic resonances X(4140), X(4274), X(4350), X(4500), and X(4700) reported by several experiment collaborations. The colour antitriplet-triplet configuration of diquark-antidiquark combinations with all conceivable quantum numbers have been used to calculate their masses. The results reveal that if the colour structure of the diquark-antidiquark configuration is $\bar{3}_{c} \otimes 3$, the tetraquarks structures may occur otherwise may observed as resonance. The X(4274) state, can be defined as the $J^{PC}$=$2^{++}$ tetraquark state, while the X(4140) state can be regarded as the $J^{PC}$=$1^{++}$ tetraquark state in this calculation. When radial excitation is considered, X(4700) may be explained as a 2S radial excited tetraquark state with $J^{PC}$=$0^{++}$. The orbitally excited states Y(4626), Y(4630) and Y(4660) can be explained as P-wave tetraquark with quantum number $1^{--}$. The masses of [$bs\bar{b}\bar{s}$] and [$bq\bar{b}\bar{s}$] are found to be in the range between 10.5 GeV- 11.5 GeV and are very close to two-meson thresholds.
\end{abstract}
%
\noindent
{\bf Program Summary and Specifications}\\
\begin{small}
\noindent
{Program title:}\\
{Licensing provisions:}\\
{Programming language:}\\
{Repository and DOI:}\\
{Description of problem:}\\
{Method of solution:}\\
{Additional comments:}\\
\end{small}
\section{Introduction}
\label{intro}
The quark model was established in 1964 to characterise mesons as bound states formed of a quark–antiquark pair and baryons as bound states made of three quarks \cite{M. Gell-Mann}. However, quantum chromodynamics (QCD) facilitates the existence of more complicated structures, which are referred to as exotic hadrons or simply exotics \cite{pdg,S. L. Olsen, R.F.Lebed}. These unique hadrons are referred to as XYZ states, and they have been identified in the quarkonium spectrum \cite{M. Cleven}. 
In the literature, many theoretical interpretations of these states have been proposed (see \cite{XinZhen,exoticstates,N Brambilla,A. Esposito,X3872,cccc,Y4260,Z3900} and references therein for current studies).  
\\
In 2009, the CDF Collaboration found the X(4140) with mass
 M = 4143.0 $\pm$ 2.9 $\pm$ 1.2 MeV and width $\Gamma$=
$11.7^{+8.4}_{-6.7} \pm 3.7$ MeV in $B^{+} \rightarrow J/\psi \phi K^{+}$ decay \cite{T. Aaltonen}.
In 2010, a narrow resonance X(4350) with mass M = 4350.6
$11.7^{+4.6}_{-5.1} \pm 0.7$ MeV and width $\Gamma$=$13^{+18}_{-9} \pm 4$ MeV was reported by the Belle
Collaboration in the $\gamma\gamma\rightarrow J/\psi \phi$
process, and the possible spin parity is $J^{PC}$ = $0^{++}$ or $2^{++}$ \cite{C.P.}. A few years later, the exotic resonance X(4100) was observed by some other collaborations including LHCb, D0, CMS, and BABAR
\cite{R. Aaij,V. M.,S. Chatrchyan}.
In 2011, another resonance X(4274) with mass M = 4274.4$ \pm 1.9$ MeV and width $\Gamma$=$32.3 \pm 7.6$ MeV was observed by the CDF Collaboration in $B^{+} \rightarrow J/\psi \phi K^{+}$ decay with 3.1$\sigma$ significance \cite{T. Aaltonen et}.
\\
\\

\begin{table*}
\centering
\caption{Experimental data on hidden-charm-strange and hidden-bottom-strange exotic states \cite{pdg}.}
\label{tab:1}
\begin{tabular}{lllll}
\hline\noalign{\smallskip}
State& $J^{PC}$  & $M$ (MeV) & Observed in & Experiment\\
\hline
X(3860) & $0^{++}$ & 3862$^{+26}_{-32}$&$e^{+}e^{-} \rightarrow J/\psi(D\bar{D})$&Belle \\
X(3872) & $1^{++}$ & 3871.69$\pm$0.17&$B^{\pm} \rightarrow J/\psi \pi^{-} \pi^{+} K^{\pm}$&Belle \\
X(3915) & $?^{??}$ & 3918.4$\pm$1.9&$e^{+}e^{-} \rightarrow J/\psi (D\bar{D}^{*})$&Belle \\
X(3940) & $?^{?+}$ & 3942$\pm$9&$e^{+}e^{-} \rightarrow J/\psi X$&Belle \\
X(4140) & $1^{++}$ & 4146.8$\pm$2.4&$B^{+} \rightarrow J/\psi \phi K^{+}$&CDF,LHCb \\
X(4160) & $?^{??}$ & 4156$\pm$15&$e^{+}e^{-} \rightarrow J/\psi (D^{*}\bar{D}^{*})$&Belle \\
X(4274) & $1^{++}$ & 4274$^{+8}_{-6}$&$B^{+} \rightarrow J/\psi \phi K^{+}$&CDF,LHCb \\
X(4350) & $?^{?+}$ & 4350$\pm$0.7&$\gamma \gamma \rightarrow \phi J/\psi$&Belle \\
X(4500) & $0^{++}$ & 4506$\pm11^{+12}_{-15}$&$B^{+} \rightarrow J/\psi \phi K^{+}$&LHCb \\
X(4700) & $0^{++}$ & 4704$\pm10^{+14}_{-24}$&$B^{+} \rightarrow J/\psi \phi K^{+}$&LHCb \\
X(4740) & $?^{?+}$ & 4741$\pm6\pm6$&$B_{s} \rightarrow J/\psi \phi \pi^{-} \pi^{+}$&LHCb \\
Y(4230) & $1^{--}$ & 4218.7$\pm$2.8&$e^{+}e^{-} \rightarrow \omega \chi_{c0}$&BESIII \\
Y(4260) & $1^{--}$ & 4230$\pm$8&$e^{+}e^{-} \rightarrow \gamma_{ISR}\pi^{-} \pi^{+}J/\psi$&BaBar \\
Y(4330) & $1^{--}$ & 4634$^{+8}_{-7}$&$e^{+}e^{-} \rightarrow \Lambda_{c}^{-} \Lambda_{c}^{+}$&Belle\\
Y(4360) & $1^{--}$ & 4368$\pm$13&$e^{+}e^{-} \rightarrow \gamma_{ISR}\pi^{-} \pi^{+}\psi(2S)$&Belle \\
Y(4390) & $1^{--}$ & 4392$\pm$7&$e^{+}e^{-} \rightarrow \pi^{-} \pi^{+}h_{c}$&BESIII \\

Y(4660) & $1^{--}$ & 4633$\pm$7&$e^{+}e^{-} \rightarrow \gamma_{ISR}\pi^{-} \pi^{+}\psi(2S)$&Belle \\
$Z_{c}(3900)^{\pm}$ & $1^{+-}$ & 3888.4$\pm$2.5&$e^{+}e^{-} \rightarrow \pi^{-} \pi^{+}J/\psi$&BESIII \\
$Z_{cs}(3985)^{\-}$ & $1^{+}$ & 3982.4$^{+8}_{-7}$ $\pm$2.1&$e^{+}e^{-} \rightarrow \pi^{-} \pi^{+}J/\psi$&BESIII \\
$Z_{c}(4020)^{\pm}$ & $1^{+-}$ & 4024.1$\pm$1.9&$e^{+}e^{-} \rightarrow \pi^{+} \pi^{-}h_{c}$&BESIII \\
$Z_{c}(4430)^{\pm}$ & $1^{+}$ & 4478$^{15}_{-18}$&$B^{\pm} \rightarrow K\pi^{\pm}\psi(2S)$&Belle \\

\noalign{\smallskip}\hline
\end{tabular}
\end{table*}

The LHCb Collaboration confirmed the existence of the X(4140) and X(4274), in the $B^{+} \rightarrow J/\psi \phi K^{+} $, decay, and their quantum numbers are measured to be $J^{PC}=1^{++}$ \cite{LHCbX4140}. In the process of $e^{-}e^{+}\rightarrow D_{s}^{+}D_{s1} (2536)^{-}$+c.c: via initial state radiation, the Belle Collaboration recently discovered a vector charmoniumlike state \cite{S. Jia}. The state has a mass $4265.9^{+6.2}_{-6.0} \pm 0.4$ and width of $49.8^{+13.9}_{-11.5} \pm 4.0$ MeV, decays into a charmed antistrange and anticharmed-strange meson pair $D_{s}^{+}D_{s1} (2536)^{-}$ with a significance of 5.9$\sigma$, which are consistent with the states Y(4630) and Y(4660) within errors despite being found in separate processes \cite{X. L.,G. Pakhlova}. The state is described as an exotic charmonium like state with $1^{--}$ named Y(4626), which provides an excellent opportunity to study the low-energy strong interaction \cite{S. Jia}. 
Then the question comes up of whether or not the main part of the states Y(4626) can be described as the tetraquark state [cs][$\bar{c}\bar{s}$]. So, the chiral constituent quark model and the quasi potential Bethe-Salpeter equation with the one-boson-exchange model were immediately used to describe the state Y(4626) as a molecular state of $D_{s}^{*}\bar{D}_{s1}(2536)$ with $1^{--}$ \cite{Y. Tan,J. He}.

The states Y(4626), Y(46300) and Y(4660)
can be uniformly described as the P-wave tetraquark state with $1^{--}$ \cite{C. Deng,Y. Yang}.

In addition to phenomenological expansions of the quark model to exotics, current non-relativistic effective field theories and lattice QCD calculations are also included in theoretical methodologies \cite{M.Yu.Barabanov,Y. R. Liu et al,Qi-Fang,C. R. Deng}.
The decay behaviour of the state Y(4626) provides the most intuitive information: the major component is most likely a tetraquark system $cs\bar{c}\bar{s}$. In the same way that the deuteron is held together by the exchange of pion and other light mesons \cite{M. Karliner2015}, Karliner and Rosner predicted the masses of tetraquark state $cs\bar{c}\bar{s}$ based on the proximity of $D_{s}\bar{D}_{s}$ pairs to thresholds \cite{M. Karliner2016}.Albuquerque and Nielsen said that the state Y(4660) was the state [cs][$\bar{c}\bar{s}$] with QCD sum rules \cite{R. M. Albuquerque}. Inspired by the states X(4140), X(4274), X(4500), X(4700), and Y(4140), the tetraquark state $cs\bar{c}\bar{s}$ was also studied in different theoretical frameworks, such as simple color-magnetic interaction models \cite{Fl Stancu,J.Wu}, the QCD sum rule \cite{W. Chen2011,H. X. Chen2017,W. Chen2017}, nonrelativistic and relativistic quark models \cite{P. G. Ortega,Y. F. Yang,Q. F. Lu}, the diquark model \cite{R. F. Lebed} and lattice QCD \cite{M. Padmanath}. 
\\
\\
Recently, an approach, which tried to unify the description of “XYZ” particles, based on the Born-Oppenheimer approximation, was applied to the tetraquark system with double heavy flavours, X(4140), X(4274), and X(4350), all taken as $cs\bar{c}\bar{s}$ tetraquarks \cite{E. Braaten}. A number of preliminary assignments have been proposed for the Y(4260), Y(4360), Y(4660), and $Z_{c}$(4020), including tetraquark states, molecular states, re-scattering effects, and so on \cite{R. N. Faustov,G. Yang,A. Hosaka}.
We may be able to discriminate between the above ideas by comparing theoretical predictions for the spectrum and principal decay modes of four-quark states to actual experimental evidence. 
\\
\\
The appearance of totally heavy $QQ\bar{Q}\bar{Q}$ bound states \cite{M. N. Anwar,A. V. Berezhnoy PRD,W. Chen,J.-M. Richard PRD} might give substantial evidence in support of the tetraquark model. The heavy-light tetraquarks in a $QQ\bar{q}\bar{q}$ configuration are equally intriguing. It would be fascinating to see whether a $QQ\bar{q}\bar{q}$ tetraquark can withstand significant decays, but there is currently no experimental evidence for this. Lipkin \cite{H. J. Lipkin} and Ader et al. \cite{J. P. Ader} were the first to demonstrate that $QQ\bar{q}\bar{q}$ is stable against significant decays theoretically. $bb\bar{q}\bar{q}$, but not its charm counterpart $cc\bar{q}\bar{q}$, nor the combined (beauty charm) $bc\bar{q}\bar{q}$ state \cite{ M. Karliner PRL,E. J. Eichten}, has recently been demonstrated to be stable despite significant decays. Refer to references \cite{A. Czarnecki,S. Q. Luo} for further information on the stability of various heavy-light tetraquarks. 
\\
Table 1 shows experimental results on exotic hidden-charm mesons. We utilise the XYZ naming scheme, with X denoting neutral exotic charmonium-like states seen in hadronic decays, Y denoting neutral exotic charmonium-like states with $J^{PC}$= $1^{--}$ seen in $e^{+}e^{-}$ collisions, and Z denoting charged (isospin triplet I = 1) charmonium-like states. The subsequent states are expressly exotic, as they cannot be $c\bar{c}$ states and, in order to have a non-zero charge, must include at least extra light quarks and antiquarks. Table 1 shows the experimentally determined quantum numbers $J^{PC}$, masses M, observation channels, and names of experiments where they were first detected. The X(4140), X(4274), X(4500), X(4700), and X(4740) states were found as resonances in the $J/\psi\phi$ mass spectrum, indicating that they should contain the strange quark and strange antiquark rather than the u and d quarks and antiquarks.
\\
\\
The following is the structure of the paper. In Section II, we discuss non-relativistic diquark model, as well as the specifics of the calculations that were performed. Section III is dedicated to a discussion of our findings as well as a comparison of our findings with those of earlier experimental and theoretical investigations. As part of this section, we also present some preliminary assignments to the XYZ states and compare them to alternative theoretical interpretations (if available). Finally, we present a brief summary of our findings.

\section{Non-relativistic Diquark-Antidiquark Model of Tetraquarks}
\label{sec:1}
The non-relativistic framework \cite{Rohitepja,debastiani
}, illustrates the interaction of constituent quarks in terms of potentials that are often inspired by phenomenology but may also be deduced via lattice QCD calculations\cite{Gyang}.
In a heavy-light quark system, a non-relativistic technique based on static potentials may be a plausible approximation since the kinetic energy of the constituents is small when compared to their rest energy. The approach starts with the solution of the time-independent radial $Schr\ddot{o}dinger$ equation in order to determine the binding energy of each specific state \cite{scr1,lucha,prd}. 
\begin{eqnarray}
\left[-\frac{1}{2\mu}\left(\frac{d^{2}}{d(r)^{2}}+\frac{2}{r}\frac{d}{d(r)}-\frac{L(L+1)}{r}^{2}\right)+ V(r)
\right]\times \nonumber \\  
\psi(r) = E\psi(r) \quad \quad  
\end{eqnarray}
where, L and E are the orbital quantum number and energy eigenvalue respectively. By substituting $\psi(r) = r^{-1}\phi(r)$ in eq.(1) modifies to;
\begin{equation}
\left[-\frac{1}{2\mu}\left(\frac{d^{2}}{dr^{2}}+\frac{L(L+1)}{r^{2}}\right)+V(r)\right]  \phi(r) = E\phi(r)  
\end{equation}
In a more simplified way we can express the Hamiltonian equation for a quark–antiquark system, as follows \cite{M. Furuichi};
\begin{eqnarray}
H\psi(r) = E\psi(r) \rightarrow (T+V(r))\psi=E\psi
\end{eqnarray} 
where T is the constituents kinetic energy and V(r) is the interaction potential. The fundamental two-body Hamiltonian center-of-mass frame of mesons and tetra quarks is denoted by the following;
\begin{equation}
H=\sum_{i=1}^{2}(M_{i}+\frac{p_{i}^{2}}{2M_{i}})+ V(r)
\end{equation}
Here $M_{i}$ is the constituent mass and $p_{i}$ is the relative momentum of the system, while $V(r)$ is the interaction potential.

%
%
A zeroth-order $V(r)$ Cornell-like potential \cite{cornell} is a reliable and extensively used potential model in the spectroscopic analysis of heavy-quarkonium systems. The Cornell-like potential $V_{C+L}(r)$ is composed of the Coulomb and linear terms, with the Coulomb part arising from a Lorentz vector exchange (basically one gluon exchange) and the linear term causing confinement typically associated with a Lorentz scalar exchange.
\begin{eqnarray}
V_{C+L}(r)=\frac{k_{s}\alpha_{s}}{r}+br\\
k_{s} = -\frac{4}{3} \quad for \quad q\bar{q}\\
= -\frac{2}{3} \quad for \quad qq \quad or \quad \bar{q}\bar{q}
\end{eqnarray}
where, $\alpha_{s}$ is known as the QCD running coupling constant, $k_ {s}$ is a colour factor, b is string tension.
The Hamiltonian may be expressed in terms of an unperturbed one-gluon exchange (OGE) potential and a relativistic mass correction term $V^{1}(r)$. We have included the relativistic mass correction term $V^{1}(r)$ originally established by Y. Coma et al. \cite{koma}, in the central potential. The final form of central potential is provided by;
\begin{equation}
 V(r) = V_{C+L}(r) + V^{1}(r)\left( \frac{1}{M_\mathcal{D}}+\frac{1}{M_\mathcal{\bar{D}}}\right)+\mathcal{O}\left(\frac{1}{m^{2}}\right)
\end{equation}
where $M_\mathcal{D}$ and $M_\mathcal{\bar{D}}$ are the masses of the constituents.
The non-perturbative form of relativistic mass correction term $V^{1}(r)$ is not yet known, but leading order perturbation theory yields \cite{koma},
\begin{equation}
V^{1}(r)=-\frac{C_{F}C_{A}}{4} \frac{\alpha^{2}_{s}}{(r)^{2}}
\end{equation}
where $C_{F}=\frac{4}{3}$ and $C_{A}=3$  are the Casimir charges of the fundamental and the adjoint representation respectively \cite{koma}.
When applied to charmonium, the relativistic mass correction is found to be equivalent to the coulombic component of the static potential, and one-fourth of the coulombic term for bottomonium. We've incorporated spin-dependent interactions in addition to the central interaction potential $V(r)$, which are perturbatively included \cite{ak,vk,dpepjc,dpepjp,dpijp1,dpfew,dpijp2}.
\\
In general, projected quark models reduce the study of quark bound states to a quantum mechanical problem, which includes the QCD dynamics. In the potential model, the spin-dependent terms are included perturbatively. This approach generates a set of four free optimal parameters that can be used to generate the meson-spectra, and later they can also be used to calculate the mass-spectra of diquarks and tetraquarks. 
\begin{table}
\caption{The fitting parameters of the model for obtaining the mass-spectra of $cs\bar{c}\bar{s}$ and $bs\bar{b}\bar{s}$. The quark masses  $m_{c}$ = 1.4 GeV, $m_{q}$ = 0.330 GeV and $m_{s}$ = 0.550 GeV have been taken from PDG \cite{pdg}.}
\label{tab:2}       
\begin{tabular}{llllllllllll}
\hline\noalign{\smallskip}
Data Set& $\alpha_{s}$ & $\sigma$ (GeV) & b (Ge$V^{2}$) \\
\noalign{\smallskip}\hline\noalign{\smallskip}
I& 0.68 & 1.10 & 0.1011  \\
II&0.66&0.8125&0.1011\\
III&   0.5167  &   1.034   &   0.1011\\
IV&   0.3901  &   1.034   &   0.1102 \\
\noalign{\smallskip}\hline
\end{tabular}
\end{table}

\begin{table*}
\centering
\caption{The Mass-Spectra of $D_{s}$ [$c\bar{s}$], generated from data set I. (Units are in MeV)}
\label{tab:3}
\footnotesize
\begin{tabular*}{170mm}{@{\extracolsep{\fill}}cccccccccccccc}
\hline
$N^{2S+1}L_{J}$&$J^{PC}$	&	$\langle E^{0}\rangle$	&	$\langle V^{(0)}_{V}\rangle$	&	$\langle V^{(0)}_{S}\rangle$	&	$\langle V^{(0)}_{SS}\rangle$	&	$\langle V^{(1)}_{LS}\rangle$	&	$\langle V^{(1)}_{T}\rangle$ &	$\langle V^{(1)}(r) \rangle $	&	$\langle K.E. \rangle$	&	$M_{f}$ 	&	 $M_{Exp}$ \cite{\cite{pdg}}&Meson\\
\hline
$1^{1}S_{0}$&$0^{-+}$	&	139	&	-486	&	260	&	-125	&	0	&	0	&	-7.4	&	492	&	1964	&1968& $D^{\pm}_{s}$ \\
$1^{3}S_{1}$	&$1^{--}$&	138	&	-486	&	260	&	41.7	&	0	&	0	&	-5.8	&	324	&	2130	&	2112 &$D^{*\pm\#\#}_{s}$\\
$2^{1}S_{0}$&$0^{-+}$	&	700	&	-265	&	559	&	-61.7	&	0	&	0	&	-4.0	&	468	&	2589	&	-&	-\\
$2^{3}S_{1}$	&$1^{--}$&	702	&	-265	&	559	&	20.5	&	0	&	0	&	-3.5	&	387	&	2673	&	2714$\pm$5 & $D^{*}_{s1}(2700)^{\pm}$ \\
$3^{1}S_{0}$&$0^{-+}$	&	1090	&	-201	&	796	&	-46.1	&	0	&	0	&	-2.8	&	541&	2993	&	-&	-\\
$3^{3}S_{1}$ &$1^{--}$	&	1090	&	-201	&	796	&	15.3	&	0	&	0	&	-2.7	&	480&	3056	&	3044& $D^{*}_{sJ}(3040)^{\pm\#\#}$\\
$4^{1}S_{0}$	&$0^{-+}$&	1414	&	-169	&	1001	&	-38.3	&	0	&	0	&	-2.0	&	620	&	3326	&	-&	-\\
$4^{3}S_{1}$&$1^{--}$	&	1415	&	-169	&	1001	&	12.7	&	0	&	0	&	-2.0	&	570	&	3378	&	-&	-\\
$1^{3}P_{0}$& $0^{++}$&	563	&	-232	&	456	&	1.8	&	-87.2	&	-40.1	&	-4.7	&	337	&	2388	&	2318 & $D^{*}_{s0}(2317)^{\pm}$\\

$1^{3}P_{1}$& $1^{++}$	&	563	&	-232	&	456	&	1.8	&	-43.7	&	20.0	&	-4.5	&	337	&	2492	&	2460 &$D_{s1}(2460)^{\pm}$\\

$1^{1}P_{1}$&$1^{+-}$	&	563	&	-232	&	456	&	-5.6	&	0	&	0	&	-4.6	&	345	&	2508	&	2535 & $D_{s1}(2536)^{\pm}$\\

$1^{3}P_{2}$&$2^{++}$	&	563	&	-232	&	456	&	1.8	&	43.6	&	-4.0	&	-3.6	&	337	&	2555	&	2569 & $D^{*}_{s2}(2573)^{\pm}$\\

$2^{3}P_{0}$&$0^{++}$	&	967	&	-175	&	705	&	2.2	&	-82.1	&	-35.7	&	-3.0	&	435	&	2802	&	-&	-\\
$2^{3}P_{1}$	&$1^{++}$ &	967	&	-175	&	705	&	2.2	&	-41.4	&	17.9	&	-3.0	&	435	&	2896	&	-&	-\\
$2^{1}P_{1}$&$1^{+-}$	&	967	&	-175	&	705	&	-6.8	&	0	&	0	&	-2.8	&	444	&	2911	&-&	-\\
$2^{3}P_{2}$&$2^{++}$	&	968	&	-175	&	705	&	2.3	&	41.4	&	-3.5	&	-2.8	&	435	&	2957	&	-&	-\\
$3^{3}P_{0}$	&$0^{++}$&	1302	&	-146	&	918	&	2.4	&	-80.1	&	-33.7	&	-2.4	&	527	&	3140	&	-&	-\\
$3^{3}P_{1}$&$1^{++}$	&	1302	&	-146	&	918	&	2.3	&	-40.2	&	16.8	&	-2.3	&	528	&	3231	&	-&	-\\
$3^{1}P_{1}$	&$1^{+-}$&	1302	&	-146	&	918	&	-7.4	&	0	&	0	&	-2.3	&	537	&	3245	&	-&	-\\
$3^{3}P_{2}$	&$2^{++}$&	1302	&	-146	&	918	&	2.4	&	40.0	&	-3.3	&	-2.2	&	527	&	3291	&	-&	-\\

$1^{3}D_{1}$&$1^{--}$	&	829	&	-164	&	609	&	0.08	&	-11.2	&	-5.1	&	-3.8	&	383	&	2762	& 2714$\pm$5	&$D^{*}_{s1}(2700)^{\pm}$\\
$1^{3}D_{2}$&$2^{--}$	&	828	&	-164	&	609	&	0.08	&	-3.7	&	5.1	&	-3.6	&	383	&	2780	&	-&-\\
$1^{1}D_{2}$&$2^{-+}$	&	829	&	-164	&	609	&	-0.2	&	0	&	0	&	-3.7	&	383	&	2778	&	-&-\\
$1^{3}D_{3}$&$3^{--}$	&	828	&	-164	&	609	&	0.08	&	7.5	&	-1.4	&	-3.7	&	383	&	2785	&	2860$\pm7$&$D^{*}_{s3}(2860)^{\pm}$\\
$2^{3}D_{1}$&$1^{--}$	&	1175	&	-136	&	831	&	0.1	&	-15.1	&	-4.9	&	-2.6	&	480	&	3105	&	-&-\\
$2^{3}D_{2}$&$2^{--}$	&	1176	&	-136	&	831	&	0.1	&	-5.1	&	4.9	&	-2.0	&	481	&	3127	&	-&-\\
$2^{1}D_{2}$&$2^{-+}$	&	1177	&	-136	&	831	&	-0.4	&	0	&	0	&	-2.4	&	482	&	3127	&	-&-\\
$2^{3}D_{3}$&$3^{--}$	&	1242	&	-136	&	831	&	0.1	&	10.1	&	-1.4	&	-2.5	&	481	&	3136	&-&-\\
[1ex]
\hline
\end{tabular*}
{$^{\#\#}$ The quantum numbers of these mesons are not assigned yet in the Recent updated PDG.}
\end{table*}

\subsection{Spin-dependent Terms}
Three spin-dependent interactions based on the Breit-Fermi Hamiltonian are introduced for one gluon exchange and are addressed using first order perturbation theory by adding their matrix components as corrections to the energy \cite{spin1,spin2}. Spin-dependent potentials $V_{SD}$, i.e. a spin-spin $V_{SS}(r)$, spin-orbit $V_{LS}(r)$, and tensor $V_{T}(r)$, are required to gain a better understanding of the splitting of orbital and radial excitations for various combinations of quantum numbers of the tetraquarks.

\begin{equation}
V_{SD} (r) = V_{SS} (r)+V_{LS} (r)+V_{T} (r),
\end{equation}
The fine structure of the states is described by the spin-orbit term $V_{LS}(r)$ and the tensor term $V_{T}(r)$, whereas the hyperfine splitting is described by the spin-spin term $V_{SS}(r)$ proportionate to 2$S_{1} \cdot S_{2}$. These spin-dependent terms of Eq. (12) may be expressed as in terms of the vector and scalar sections of the static potential V(r).
\begin{equation}
V_{SS} (r) = \frac{2}{3M_{\mathcal{D}} M_{\mathcal{\bar{D}}}} \nabla^2 V_{V}(r)S_{1} \cdot S_{2} = -\frac{8k_{s}\alpha_{s}\pi}{3M_{\mathcal{D}} M_{\mathcal{\bar{D}}}} \delta^3(r)  S_{1} \cdot S_{2},
\end{equation}
In heavy quarkonium spectroscopy, a reasonable agreement may be obtained by incorporating spin-spin interactions in zero-order potentials using the $Schr\ddot{o}dinger$ equation by including the spin-spin interaction using the artefact introducing a new parameter $\sigma$ instead of the Dirac delta \cite{griffiths}. 
As a result, $V_{SS}(r)$ may now be redefined as follows:
\begin{equation}
V_{SS} (r) = -\frac{8 \pi k_{s}\alpha_{s}}{3M_{\mathcal{D}} M_{\mathcal{\bar{D}}}} (\frac{\sigma}{\sqrt{\pi}})^{3} \exp^{-\sigma^2 (r)^2} S_{1} \cdot S_{2},
\end{equation} 
The spin-orbit interaction term can be defined as;
\begin{eqnarray}
V_{LS} (r)= \frac{1}{2M_{\mathcal{D}} M_{\mathcal{\bar{D}}}}\frac{1}{r} [3\frac{dV_{V}(r)}{dr}-\frac{dV_{S}}{dr}] L \cdot S\\
 = [-\frac{3k_{s}\alpha_{s}\pi}{2M_{\mathcal{D}} M_{\mathcal{\bar{D}}}}\frac{1}{(r)^{3}}-\frac{b}{2M_{\mathcal{D}} M_{\mathcal{\bar{D}}}}\frac{1}{(r)}] L \cdot S,
\end{eqnarray}
Furthermore, the tensor term can be derived and utilised in our calculation;
\begin{equation}
V_{T}(r) = C_{T}(r)\left( \frac{(S_{1}\cdot (r))(S_{2}\cdot (r))}{(r)^2}- \frac{1}{3} (S_{1} \cdot S_ {2})\right) 
\end{equation}
where;
\begin{equation}
C_{T}(r) = -\frac{12k_{s}\alpha_{s}\pi}{4m^2}\frac{1}{(r)^{3}}
\end{equation}
The results of $(S_{1} \cdot S_ {2})$ may be obtained by solving the diagonal matrix elements for the spin-$\frac{1}{2}$ and spin-1 particles, as detailed in the following Refs. \cite{debastiani,prd}. To solve the Equ.17 in a more simplified form the equation may be expressed as;
\begin{equation}
\mathbf{{S_{12}}}=12\left(\mathbf{ \frac{(S_{1}\cdot (r))(S_{2}\cdot (r))}{(r)^2}- \frac{1}{3} (S_{1} \cdot S_ {2})}\right) 
\end{equation} 
and which can be redefined as ;
\begin{equation}
\mathbf{{S_{12}}}= 4\left[3\mathbf{(S_{1}\cdot \hat{(r)})(S_{2}\cdot \hat{(r)})-(S_{1} \cdot S_ {2})} \right]
\end{equation}
Pauli matrices and spherical harmonics with their corresponding eigenvalues may be used to achieve the results of the $S_{12}$ term. The following conclusions are valid for bottomonium and diquarks \cite{bethe,thesis} ;
\begin{align}
\mathbf{\left\langle S_{12}\right\rangle}_{\frac{1}{2}\otimes\frac{1}{2}\rightarrow S=1, l\neq0} = -\frac{2l}{2l+3}, for J= l+1,\\
= -\frac{2(l+1)}{(2l-1)}, for J= l-1, and\\
= +2, for J=l
\end{align}
When $l=0$ and $S=0$ the $\mathbf{\left\langle S_{12}\right\rangle}$ always vanishes, but it yields a non-zero value for excited states in mesons: $\mathbf{\left\langle S_{12}\right\rangle} = -\frac{2}{5}, +2, -4$ for $J=2,1,0$, respectively.
These values are valid only for bottomonium and diquarks that are specifically spin-half particles, but in the case of tetraquarks when spin-1 diquarks are involved, it needs a laborious algebra, which is not discussed in depth here, rather one can refer Refs. \cite{bethe,thesis} for detailed discussion. The results for tensor interaction of tetraquarks will obtained by the same formula which is used in case of bottomonium except that the wavefunction obtained here will be of spin 1 (anti)diquark.
\begin{align}
\mathbf{S_{{d}-{\bar{d}}}}=12\left(\mathbf{\frac{(S_{d}\cdot (r))(S_{\bar{d}}\cdot (r))}{(r)^2}- \frac{1}{3} (S_{d} \cdot S_ {\bar{d}})}\right)\\
= \mathbf{S_{14} + S_{13} + S_{24} + S_{23}}
\end{align}
where $S_{d}$ is the total spin of the diquark, $S_{\bar{d}}$ is the total spin of the antidiquark. When the two-body problem is solved to obtain the masses of the tetraquarks, the interaction between the two (anti) quarks inside the (anti) diquark is identical; because (anti) diquarks are only considered in the S-wave state, only the spin-spin interaction is relevant; the spin-orbit and tensor are both identically zero.
Because the tetraquark radial wavefunction is obtained by treating the diquark and antidiquark as two body problem, it is reasonable to assume that the radial-dependence of the tensor term is the same for these four [$q\bar{q}$] interactions and can be obtained using the radial wavefunction with Eq (14). The following functional form for spin $\frac{1}{2}$ particles does not use any specific relation or eigenvalues, instead relying on general angular momentum elementary theory. Within this approximation, generalization of tensor operator can be consider a sum of four tensor interaction between four quark-antiquark pair as illustrated in \cite{thesis}.
\\
\\
A thorough discussion on tensor interaction can be found in Ref. \cite{thesis}.All spin-dependent terms have been computed for spin-1 diquarks and antidiquarks that combine to produce a color singlet tetraquark with spin $S_{T}$ = 0,1,2. The mass-spectra of radial and orbital excitations are obtained by coupling the total spin $S_{T}$ with the total orbital angular momentum $L_{T}$, which results in the total angular momentum $J_{T}$.
To obtain the quantum numbers ($J^{PC}$) of the tetra-quark states, one can use the following formula; $P_{T}=(-1)^{L_{T}}$ and $C_{T}=(-1)^{L_{T} + S_{T}}$.

\begin{table*}
\centering
\caption{The Mass-Spectra of B-mesons [$b\bar{s}$], generated from data set II. (Units are in MeV)}
\label{tab:4}
\footnotesize
\begin{tabular*}{170mm}{@{\extracolsep{\fill}}ccccccccccccc}
\hline
$N^{2S+1}L_{J}$	& $J^{PC}$&	$\langle E\rangle$	&	$\langle V_{V}\rangle$	&	$\langle V_{S}\rangle$	&	$\langle V_{SS}\rangle$	&	$\langle V_{LS}\rangle$	&	$\langle V_{T}\rangle$ &	$\langle V^{(1)}(r) \rangle $	&	$\langle K.E. \rangle$	&	$M_{f}$  & $M_{Exp}$ \cite{pdg}& Meson \\
\hline
$1^{1}S_{0}$&$0^{-+}$	&	69	&	-542	&	231	&	-35	&	0	&	0	&	-6.5	&	415	&	5367&5366.88&$B^{0}_{s}$\\
$1^{3}S_{1}$&$1^{--}$	&	69	&	-542	&	231	&	11	&	0	&	0	&	-5.9	&	369	&	5414&5415$^{+1.8}_{-1.5}$&$B^{*}_{s}$\\
$2^{1}S_{0}$&$0^{-+}$	&	622	&	-285	&	513	&	-13	&	0	&	0	&	-3.6	&	407	&	5941&-&-\\
$2^{3}S_{1}$	&$1^{--}$&	624	&	-285	&	513	&	4.6	&	0	&	0	&	-3.4	&	391	&	5962&-&-\\
$3^{1}S_{0}$&$0^{-+}$	&	993	&	-215	&	736	&	-9.4	&	0	&	0	&	-2.2	&	483	&	6318&-&-\\
$3^{3}S_{1}$	&$1^{--}$&	995	&	-215	&	736	&	3.1	&	0	&	0	&	-2.3	&	470	&	6331&-&-\\
%

$1^{3}P_{0}$&$0^{++}$	&	500	&	-250	&	418	&	1.1	&	-36	&	-15	&	-4.3	&	330	&	5783&-&-\\
$1^{3}P_{1}$&$1^{++}$	&	498	&	-250	&	418	&	1.1	&	-17	&	7.7	&	-4.3	&	328	&	5822&5853$\pm$15&$B^{*}_{sJ}(5850)$\\
$1^{1}P_{1}$&$1^{+-}$	&	498	&	-250	&	418	&	-3.3	&	0	&	0	&	-4.3	&	333	&	5828&5828.70&$B^{0}_{s1}(5830)^{0}$\\
$1^{3}P_{2}$&$2^{++}$	&	498	&	-250	&	418	&	1.1	&	17	&	-1.5	&	-4.3	&	328	&	5849&5839.66&$B^{*}_{s2}(5840)^{0}$\\
$2^{3}P_{0}$&$0^{++}$	&	881	&	-187	&	653	&	0.9	&	-32	&	-13.6	&	-2.7	&	414	&	6169&-&-\\
$2^{3}P_{1}$&$1^{++}$	&	883	&	-187	&	653	&	0.9	&	-16	&	6.0	&	-2.7	&	416	&	6208&-&-\\
$2^{1}P_{1}$&$1^{+-}$	&	883	&	-187	&	653	&	-3.4	&	0	&	0	&	-2.8	&	421	&	6213&-&-\\
$2^{3}P_{2}$&$2^{++}$	&	883	&	-187	&	653	&	0.9	&	-16	&	-1.3	&	-2.6	&	416	&	6199&-&-\\

$1^{3}D_{1}$&$1^{--}$	&	755	&	-175	&	564	&	0.1	&	-6.0	&	-2.0	&	-3.4	&	366	&	6080&-&-\\
$1^{3}D_{2}$&$2^{--}$	&	755	&	-175	&	564	&	0.1	&	-2.0	&	1.9	&	-3.2	&	366	&	6088&-&-\\
$1^{1}D_{2}$&$2^{-+}$	&	755	&	-175	&	564	&	-0.3	&	0	&	0	&	-3.1	&	366	&	6087&-&-\\
$1^{3}D_{3}$&$3^{--}$	&	755	&	-175	&	564	&	0.1	&	4.0	&	-0.5	&	-3.0	&	366	&	6091&-&-\\
[1ex]
\hline
\end{tabular*}
{$^{\#\#}$ The quantum numbers of these mesons are not assigned yet in the Recent updated PDG.}
\end{table*}

\section{Numerical calculation and Discussion }
All the fitting parameters of the model are fixed from the consideration of meson-spectra and this methodology is inspired from previous studies. In the present work there are four fitting parameters (m, $\alpha_{s}$, b, $\sigma$) for which the model mass ($M_{i}^{f}$) of the particular tetraquark states have been calculated. 
\\
The set of parameters \cite{pdg,griffiths} are varied in the range of ;

{\centering
0.05 $\leq$ $\alpha_{s}$ $\leq$ 0.70\\
 0.01 Ge$V^2$ $\leq$ b $\leq$ 0.450 Ge$V^2$\\
 0.05 GeV $\leq$ $\sigma$ $\leq$ 1.50 GeV\\
 4.00 GeV $\leq$ $m_{b}$ $\leq$ 5.00 GeV\\
 0.3 GeV $\leq$ $m_{q}$ $\leq$ 0.350 GeV\\ }
The mass-spectra of charm-strange $D_{s}$ and charm-strange (anti)diquark have been obtained from data set I. Similarly, the mass-spectra of $B_{s}$-mesons and bottom-strange (anti)diquark obtained from data set II. The mass-spectra of $cs\bar{c}\bar{s}$ and $cq\bar{c}\bar{s}$ tetraquarks are computed using data set III. The mass-spectra of $bs\bar{b}\bar{s}$ and $bq\bar{b}\bar{s}$ tetraquarks are computed using data set IV.


\subsection{$D_{s}$ and $B_{s}$ mesons spectrum}
\label{sec:2}
To calculate the mass-spectra of diquarks and tetra-quarks, first, we estimate the mass-spectra of quarkonium states [$c\bar{q}$] and [$c\bar{s}$] whose results are tabulated in Table \ref{tab:2} and Table \ref{tab:3}, respectively. The model's reliability and consistency have been tested by obtaining the mass-spectra of heavy and heavy-light bottom mesons.
The SU(3) color symmetry allows only colorless quark combination $|Q\bar{Q}\rangle$ to form any color  singlet state \cite{griffiths,thesis}, as in our case [$c\bar{q}$] and [$c\bar{s}$] are mesons and exhibits $|Q\bar{Q}\rangle: \mathbf{3 \otimes\bar{3}=1\oplus 8}$ representation which leads to carry a color factor $k_{s}=-\frac{4}{3}$ \cite{debastiani}.
The masses of the particular [$c\bar{q}$] and [$c\bar{s}$] states are obtained namely
\begin{equation} 
M_{(b\bar{s})} = M_{b}+ M_{\bar{s}}+ E_{b\bar{s}} + \langle V^{1}(r)\rangle
\end{equation}
and
\begin{equation} 
M_{(c\bar{s})} = M_{c}+ M_{\bar{s}}+ E_{c\bar{s}} + \langle V^{1}(r)\rangle
\end{equation}
The final masses obtained from the above expression constitute the contributions from different spin-dependent terms (spin-spin, spin-orbital and tensor) have tabulated in Table \ref{tab:2} and Table \ref{tab:3}.
The mass-spectra of the mesons produced in this study are compatible with the experimental data available in the most recent updated PDG \cite{pdg}. Additionally, the current work's findings are consistent with those in Ref. \cite{prd}, where the author computed the mass-spectra of heavy-light tetraquarks [$Qq\bar{Q}\bar{q}$] (Q = b, c and q = u, d) including all heavy tetraquarks.
\\
\\
There are a total of 11 charmed mesons ($c\bar{q}$) produced from the model, all of which have masses fairly closed to those predicted experimentally. In the case of   S-wave charmed mesons states the discrepancy ($\Delta$) is around 30 MeV. Particularly, in second radial states where the strength of spin-spin interaction declines sharply which leads to maximizes $\Delta$.
Similarly, there are $D_{s}$-mesons whose masses have also good agreement with experimentally predicted data and have discrepancy ($\Delta$) nearly 50-60 MeV. 
At high energy scale, discrepancy nearly 30-60 MeV's between the model's mass and experimental data can be tolerated and the fitting parameters are assumed as best fit. 
\\
\\
Spin-dependent interactions are crucial in heavy quarkonium study specially in quarkonium because they allow for the consideration of QCD dynamics in the heavy quark scenario, which lies between the perturbative and non-perturbative regimes \cite{thesis}. The spin-spin interaction's involvement in orbitally excited states is particularly intriguing. We notice that the masses of P-wave states are more precise than those of S-wave states owing to spin-orbit and tensor contributions (see Table \ref{tab:2} and Table \ref{tab:3}). The slighter contribution of the relativistic term, which is greater for lighter quarks, alters the spectrum by a few MeV's.

\begin{table*}
\centering
\caption{The masses scalar (anti)diquarks and energy eigenvalue from present work and comparison with other prior works. (Units are in MeV).}
\label{tab:5}       
\begin{tabular}{llllllllll}
\hline
Diquark &$\langle E^{0}\rangle$ & Ours &\cite{XinZhen}&\cite{R. N. Faustov}& \cite{mb}& \cite{mh} & \cite{zg}\\
\hline
 cs & 192 & 2119 &2138& 2091&2229&2091& 1980 $\pm$0.10 &&\\
 bs & 160 & 5496 &5494& 5462&5572&5462&5350$\pm$0.09 &&\\
\noalign{\smallskip}\hline
\end{tabular}
\end{table*}
\subsubsection{Diquarks}
A (anti)diquark ($\mathcal{(\bar{D})D}$)is a pair of (anti)quarks that interact with one another through gluonic exchange and can form a bound state \cite{exoticstates,N Brambilla,A. Esposito,Rohitepja}. It's important to remember that while constructing a diquark ($\mathcal{D}$) is a composite (qq) system not a point-like object. The form factor, which can be represented as the overlap integral of diquark wave functions, does indeed smear its interaction with gluons. The Pauli principle should also be considered, which results in the following ground state diquark limitations.
To comply the Pauli exclusion principle, which states that diquarks with the same flavour quark should have a spin of 1, the diquark's wavefunction must be antisymmetric \cite{griffiths}. The (qq') diquark, which is made of quarks of various flavours, may have spins S = 0,1 (scalar [qq'], axial vector $\lbrace qq'\rbrace$ diquarks), while the $\lbrace qq\rbrace$ diquark, which is composed of quarks of the same flavour, can only have spin S = 1. Because of the stronger attraction owing to the spin–spin interaction, the scalar S diquark is frequently referred to as a ``good" diquark, while the heavier axial vector diquark is referred to as a ``bad" diquark \cite{exoticstates}.
\\
\\
To produce the most compact diquark, we will utilise the ground state ($1^{1}S_{0}$) diquarks [$cq$] or [$cs$], which have no orbital or radial excitations.
%
Everything done for quark-quark interactions is assumed to be identical to antiquark-antiquark interactions, with the exception that colors are replaced by anticolors. We utilised the same methodology as in the case of the heavy-light mesons to get the mass-spectra of (anti)diquarks.
According to QCD color symmetry, two quarks are combined in the fundamental ({\bf3}) representation to obtain the diquark, presented by  $\mathbf{3 \otimes 3 = \bar{3} \oplus 6}$.
Moreover, antiquarks are combined in the $\mathbf{\bar{3}}$ representation and can be presented as $\mathbf{\bar{3} \otimes \bar{3} = 3 \oplus \bar{6}}$ \cite{debastiani,griffiths}. 
The diquark-antidiquark approximation is significant because it reduces a complex four-body problem to a simple two-body problem \cite{prd,mh}. The hamiltonian, on the other hand, ceases replicating the meson spectra when doing the full four-body  basis treatment \cite{Gyang}. The explanation for this is simple: the $3 \otimes \bar{3}$ color coupling can be transformed into a $1\otimes1$ state, and also a $8\otimes8$ state.
The QCD color symmetry  produces a color factor $k_{s}=-\frac{2}{3}$ in antitriplet state and makes the short distance part $(\frac{1}{r})$ of the interaction attractive \cite{griffiths,thesis}.
\\
\\
We compared the diquark masses acquired in this work to those obtained in the other prior investigations mentioned in Table \ref{tab:4}. The diqaurk mass of [cq] have been taken from our previous work \cite{Rohitepja}.
The diquark masses are investigated in Ref. \cite{mb} using using a relativized diquark model Hamiltonian, which account for kinetic energy as well as splitting in the spin-spin, spin-orbit, and tensor interactions.
The masses of diquarks calculated in this research are consistent with \cite{R. N. Faustov,mh} and are less than those reported in Ref. \cite{mb}, whereas models based on QCD sum rules, such as presented in Ref \cite{zg} predict smaller diquark masses. The discrepancies may be due to the addition of new and updated data in this study.

\begin{table*}[h]
\centering
\caption{The Mass-Spectra of [$cs\bar{c}\bar{s}$], generated from data set III. (Units are in MeV)}
\label{tab:6}
\footnotesize
\begin{tabular*}{170mm}{@{\extracolsep{\fill}}ccccccccccccc}
\hline
 $N^{2S+1}L_{J}$&$J^{PC}$	&	$\langle E^{0}\rangle$	&	$\langle V^{(0)}_{V}\rangle$	&	$\langle V^{(0)}_{S}\rangle$	&	$\langle V^{(0)}_{SS}\rangle$	&	$\langle V^{(1)}_{LS}\rangle$	&	$\langle V^{(1)}_{T}\rangle$ &	$\langle V^{(1)}(r) \rangle $	&	$\langle K.E. \rangle$	&	$M^{i}_{f}$ & $M_{th}$ &Threshold\\
\hline
$1^{1}S_{0}$	&	$0^{++}$	&	-77	&	-625	&	160	&	-114	&	0	&	0	&	-3.1	&	503	&	4045&3936&$D^{\pm}_{s}D^{\pm}_{s}$\\

$1^{3}S_{1}$	&	$1^{+-}$	&	-77	&	-625	&	160	&	-57	&	0	&	0	&	-3.0	&	446	&	4103& 4080&$D^{\pm}_{s}D^{*\pm}_{s}$\\
$1^{3}S_{1}$	&	$1^{++}$&	-77	&	-625	&	160	&	0	&	0	&	0	&	-3.0	&	388	&	4161&4080&$D^{\pm}_{s}D^{*\pm}_{s}$\\
$1^{5}S_{2}$	&	$2^{++}$	&	-77	&	-625	&	160	&	57	&	0	&	0	&	-3.0	&	331	&	4218&4224&$D^{*\pm}_{s}D^{*\pm}_{s}$\\
$2^{1}S_{0}$	&	$0^{++}$	&	418	&	-300	&	380	&	-36	&	0	&	0	&	-1.8	&	374	&	4620&4682&$D^{\pm}_{s}D^{*\pm}_{s1}$\\
$2^{3}S_{1}$	&	$1^{+-}$	&	418	&	-300	&	380	&	-18	&	0	&	0	&	-1.8	&	355	&	4638&...&...\\
$2^{3}S_{1}$&	$1^{++}$	&	418	&	-300	&	380	&	0	&	0	&	0	&	-1.6	&	337	&	4656&...&...\\
$2^{5}S_{2}$	&	$2^{++}$	&	418	&	-300	&	380	&	18	&	0	&	0	&	-1.6	&	319	&	4675&4682&$D^{\pm}_{s}D^{*\pm}_{s1}$\\
$3^{1}S_{0}$	&	$0^{++}$	&	722	&	-220	&	556	&	-23	&	0	&	0	&	-0.9	&	410	&	4848&...&...\\
$3^{3}S_{1}$	&	$1^{+-}$	&	722	&	-220	&	556	&	-11	&	0	&	0	&	-0.9	&	399	&	4949&...&...\\
$3^{3}S_{1}$&	$1^{++}$	&	722	&	-220	&	556	&	0	&	0	&	0	&	-0.9	&	387	&	4960&...&...\\
$3^{5}S_{2}$	&	$2^{++}$	&	722	&	-220	&	556	&	11	&	0	&	0	&	-0.9	&	375	&	4972&5012&$D^{\pm}_{s}D^{*}_{sJ}$\\

$1^{1}P_{1}$	&	$1^{--}$	&	330	&	-265	&	309	&	-11	&	0	&	0	&	-2.2	&	296	&	4556&4647&$D^{*\pm} D^{\pm}_{s1}$\\
$1^{3}P_{0}$	&	$0^{-+}$	&	330	&	-265	&	309	&	-6.0	&	-46	&	-36	&	-2.1	&	374	&	4477&4286&$D^{\pm}_{s}D^{*\pm}_{s0}$\\
$1^{3}P_{1}$	&	$1^{-+}$	&	330	&	-265	&	309	&	-6.0	&	-23	&	18	&	-2.2	&	295	&	4557&4428&$D^{\pm}_{s}D^{\pm}_{s1}$\\
$1^{3}P_{2}$	&	$2^{-+}$	&	330	&	-265	&	309	&	-6.0	&	23	&	-4.0&	-2.2	&	270	&	4581&4537&$D^{\pm}_{s}D^{*\pm}_{s2}$\\
$1^{5}P_{1}$	&	$1^{--}$	&	330	&	-265	&	309	&	6.0	&	-70	&	-25	&	-2.0	&	375	&	4476&4647&$D^{*\pm} D^{\pm}_{s1}$\\
$1^{5}P_{2}$	&	$2^{--}$	&	330	&	-265	&	309	&	6.0	&	-23	&	25	&	-2.0	&	277	&	4575&4572&$D^{*\pm} D^{\pm}_{s1}$\\
$1^{5}P_{3}$	&	$3^{--}$	&	330	&	-265	&	309	&	6.0	&	46	&	-7.0	&	-2.2	&	240	&	4612&4685&$D^{*\pm} D^{*\pm}_{s2}$\\

$2^{1}P_{1}$	&	$1^{--}$	&	641	&	-194	&	493	&	-10	&	0	&	0	&	-1.3	&	353	&	4869&...&...\\
$2^{3}P_{0}$	&	$0^{-+}$	&	641	&	-194	&	493	&	-5.0	&	-39	&	-30	&	-1.3	&	418	&	4804&...&...\\
$2^{3}P_{1}$	&	$1^{-+}$	&	641	&	-194	&	493	&	-5.0	&	-19	&	15	&	-1.3	&	352	&	4870&...&...\\
$2^{3}P_{2}$	&	$2^{-+}$	&	641	&	-194	&	493	&	-5.0	&	19	&	-3.0	&	-1.3	&	331	&	4892&...&...\\
$2^{5}P_{1}$	&	$1^{--}$	&	641	&	-194	&	493	&	5.0	&	-60	&	-21	&	-1.3	&	418	&	4804&...&...\\
$2^{5}P_{2}$	&	$2^{--}$	&	641	&	-194	&	493	&	5.0	&	-19	&	21	&	-1.3	&	335	&	4887&...&...\\
$2^{5}P_{3}$	&	$3^{--}$	&	641	&	-194	&	493	&	5.0	&	39	&	-6.0	&	-1.3	&	303	&	4919&...&...\\
[1ex]
\hline
\end{tabular*}
\end{table*}

\begin{table*}[h]
\centering
\caption{The Mass-Spectra of [$cq\bar{c}\bar{s}$], generated from data set III. (Units are in MeV)}
\label{tab:7}
\footnotesize
\begin{tabular*}{170mm}{@{\extracolsep{\fill}}ccccccccccccc}
\hline
 $N^{2S+1}L_{J}$&$J^{PC}$	&	$\langle E^{0}\rangle$	&	$\langle V^{(0)}_{V}\rangle$	&	$\langle V^{(0)}_{S}\rangle$	&	$\langle V^{(0)}_{SS}\rangle$	&	$\langle V^{(1)}_{LS}\rangle$	&	$\langle V^{(1)}_{T}\rangle$ &	$\langle V^{(1)}(r) \rangle $	&	$\langle K.E. \rangle$	&	$M^{i}_{f}$ & $M_{th}$ &Threshold\\
\hline
$1^{1}S_{0}$	&$0^{++}$&	-60	&	-608	&	165	&	-66	&	0	&	0	&	-3.7	&	449	&	3955&3837&$D^{\pm}D^{\pm}_{s}$\\
$1^{3}S_{1}$	&$1^{+-}$&	-60	&	-608	&	165	&	-33	&	0	&	0	&	-3.6	&	416	&	3988&3981&$D^{\pm}D^{*\pm}_{s}$ \\
$1^{3}S_{1}$	&$1^{++}$&	-60	&	-608	&	165	&	0	&	0	&	0	&	-3.6	&	383	&	4021&3981&$D^{\pm}D^{*\pm}_{s}$ \\
$1^{5}S_{2}$	&$2^{++}$&	-60	&	-608	&	165	&	33	&	0	&	0	&	-3.4	&	350	&	4054&4122&$D^{*\pm}D^{*\pm}_{s}$\\
$2^{1}S_{0}$	&$0^{++}$&	436	&	-294	&	390	&	-18	&	0	&	0	&	-1.7	&	358	&	4500&4518&$D^{0}_{0}D^{\pm}_{s}$\\
$2^{3}S_{1}$	&$1^{+-}$&	436	&	-294	&	390	&	-9.0	&	0	&	0			&	-1.5	&	350	&	4509&4569&$D^{\pm}D^{*\pm}_{s1}$\\

$2^{3}S_{1}$	&$1^{++}$&	436	&	-294	&	390	&	0	&	0	&	0			&	-1.5	&	340	&	4518&4569&$D^{\pm}D^{*\pm}_{s1}$\\
$2^{5}S_{2}$	&$2^{++}$&	436	&	-294	&	390	&	9.0	&	0	&	0			&	-1.3	&	332	&	4527&4608&$D^{*\pm}D^{\pm}_{s}$ \\
$3^{1}S_{0}$	&$0^{++}$&	743	&	-218	&	570	&	-11	&	0	&	0			&	-1.1	&	404	&	4814&-&-\\
$3^{3}S_{1}$	&$1^{+-}$&	743	&	-218	&	570	&	-5.8	&	0	&	0			&	-1.1	&	398	&	4820&4909&$D^{\pm}D^{\pm}_{sJ}$\\
$3^{3}S_{1}$	&$1^{++}$&	743	&	-218	&	570	&	0	&	0	&	0			&	-1.1	&	392	&	4825&4909&$D^{\pm}D^{\pm}_{sJ}$\\
$3^{5}S_{2}$	&$2^{++}$&	743	&	-218	&	570	&	5.8	&	0	&	0			&	-1.1	&	387	&	4832&-&-\\	
$1^{1}P_{1}$	&$1^{--}$&	344	&	-260	&	317	&	-12.7	&	0	&	0	&	-2.2	&	299	&	4413&4388&$D_{1}(2420)D^{\pm}_{s}(1968)$\\
$1^{3}P_{0}$	&$0^{-+}$&	344	&	-260	&	317	&	-6.3	&	-47.4	&	-38	&	-2.2	&	378	&	4334&4329&$D^{\pm}(1869)D^{\pm}_{s}(2460)$\\
$1^{3}P_{1}$	&$1^{-+}$&	344	&	-260	&	317	&	-6.3	&	-23.4	&	19	&	-2.2	&	298	&	4414&4404&$D^{\pm}(1869)D^{\pm}_{s1}(2536)$\\
$1^{3}P_{2}$	&$2^{-+}$&	344	&	-260	&	317	&	-6.3	&	23.7	&	-3.7	&	-2.2	&	273	&	4439&4428&$D^{*\pm}(2460)D^{\pm}_{s}(1968)$\\
$1^{5}P_{1}$	&$1^{--}$&	344	&	-260	&	317	&	6.3	&	-71.4	&	-26	&	-2.2	&	378	&	4334&4328&$D^{*}(2010)^{\pm} D^{*}_{s0}(2317)$\\
$1^{5}P_{2}$	&$2^{--}$&	344	&	-260	&	317	&	6.3	&	-23.7	&	26	&	-2.2	&	278	&	4435&4438&$D^{\pm}(1869) D^{*}_{s2}(2573)^{\pm}$\\
$1^{5}P_{3}$	&$3^{--}$&	344	&	-260	&	317	&	6.3	&	47.4	&	-7.5	&	-2.2	&	240	&	4472&4473&$D^{*}_{0}(2300) D^{*\pm}_{s}(2112)$\\
$2^{1}P_{1}$	&$1^{--}$&	661	&	-191	&	505	&	-9.8	&	0	&	0	&	-1.3	&	356	&	4733&-&-\\
$2^{3}P_{0}$	&$0^{-+}$&	661	&	-191	&	505	&	-4.9	&	-41	&	-32	&	-1.2	&	424	&	4666&-&-\\
$2^{3}P_{1}$	&$1^{-+}$&	661	&	-191	&	505	&	-4.9	&	-20	&	16	&	-1.4	&	356	&	4734&-&-\\
$2^{3}P_{2}$	&$2^{-+}$&	661	&	-191	&	505	&	-4.9	&	20.3	&	-3.1	&	-1.2	&	335	&	4756&-&-\\
$2^{5}P_{1}$	&$1^{--}$&	661	&	-191	&	505	&	4.9	&	-61	&	-22.2	&	-1.2	&	426	&	4665&-&-\\
$2^{5}P_{2}$	&$2^{--}$&	661	&	-191	&	505	&	4.9	&	-20	&	22	&	-1.2	&	340	&	4750&-&-\\
$2^{5}P_{3}$	&$3^{--}$&	661	&	-191	&	505	&	4.9	&	40	&	-6.3	&	-1.2	&	308	&	4783&-&-\\
[1ex]
\hline
\end{tabular*}
\end{table*}

\begin{table*}[h]
\centering
\caption{The Mass-Spectra of [$bs\bar{b}\bar{s}$], generated from data set IV. (Units are in MeV)}
\label{tab:8}
\footnotesize
\begin{tabular*}{170mm}{@{\extracolsep{\fill}}ccccccccccccc}
\hline
 $N^{2S+1}L_{J}$&$J^{PC}$	&	$\langle E^{0}\rangle$	&	$\langle V^{(0)}_{V}\rangle$	&	$\langle V^{(0)}_{S}\rangle$	&	$\langle V^{(0)}_{SS}\rangle$	&	$\langle V^{(1)}_{LS}\rangle$	&	$\langle V^{(1)}_{T}\rangle$ &	$\langle V^{(1)}(r) \rangle $	&	$\langle K.E. \rangle$	&	$M^{i}_{f}$ & $M_{th}$ &Threshold\\
\hline
$1^{1}S_{0}$	&	$0^{++}$	&	-240	&	-806	&	110	&	-40	&	0	&	0	&	-2.2	&	496	&	10711&10732& $B^{0}_{s}B^{0}_{s}$\\
$1^{3}S_{1}$	&	$1^{+-}$	&	-240	&	-806	&	110	&	-20	&	0	&	0	&	-2.0	&	476	&	10731&10781&$B^{0}_{s}B^{*}_{s}$\\
$1^{3}S_{1}$	&	$1^{++}$	&	-240	&	-806	&	110	&	0	&	0	&	0	&	-2.0	&	456	&	10751&10781&$B^{0}_{s}B^{*}_{s}$\\
$1^{5}S_{2}$	&	$2^{++}$	&	-240	&	-806	&	110	&	20	&	0	&	0	&	-2.0	&	435	&	10772&10830&$B^{*}_{s}B^{*}_{s}$\\
$2^{1}S_{0}$	&	$0^{++}$	&	261	&	-341	&	289	&	-9.0	&	0	&	0	&	-0.9	&	322	&	11244&-&-\\
$2^{3}S_{1}$	&	$1^{+-}$	&	261	&	-341	&	289	&	-4.5	&	0	&	0	&	-0.8	&	319	&	11250&-&-\\
$2^{3}S_{1}$	&	$1^{++}$	&	261	&	-341	&	289	&	0	&	0	&	0	&	-0.8	&	313	&	11254&-&-\\
$2^{5}S_{2}$	&	$2^{++}$	&	261	&	-341	&	289	&	4.5	&	0	&	0	&	-0.9	&	309	&	11259&-&-\\
$3^{1}S_{0}$	&	$0^{++}$	&	532	&	-244	&	436	&	-5.6	&	0	&	0	&	-0.6	&	345	&	11518&-&-\\
$3^{3}S_{1}$	&	$1^{+-}$	&	532	&	-244	&	436	&	-2.8	&	0	&	0	&	-0.6	&	342	&	11521&-&-\\
$3^{3}S_{1}$	&	$1^{++}$	&	532	&	-244	&	436	&	0	&	0	&	0	&	-0.6	&	339	&	11524&-&-\\
$3^{5}S_{2}$	&	$2^{++}$	&	532	&	-244	&	436	&	2.8	&	0	&	0	&	-0.6	&	337	&	11527&-&-\\
$1^{1}P_{1}$	&	$1^{--}$	&	198	&	-308	&	235	&	-5.3	&	0	&	0	&	-1.0	&	276	&	11184&-&-\\
$1^{3}P_{0}$	&	$0^{-+}$	&	198	&	-308	&	235	&	-2.6	&	-22	&	-16	&	-1.0	&	312	&	11149&-&-\\
$1^{3}P_{1}$	&	$1^{-+}$	&	198	&	-308	&	235	&	-2.6	&	-11	&	8.1	&	-1.0	&	276	&	11184&11196&$B^{0}_{s}B^{0}_{s1}(5830)$\\
$1^{3}P_{2}$	&	$2^{-+}$	&	198	&	-308	&	235	&	-2.6	&	11	&	-1.6	&	-1.0	&	264	&	11196&11206&$B^{0}_{s}B^{*}_{s2}(5840)$\\
$1^{5}P_{1}$	&	$1^{--}$	&	198	&	-308	&	235	&	2.6	&	-32	&	-11	&	-1.0	&	312	&	11148&-&-\\
$1^{5}P_{2}$	&	$2^{--}$	&	198	&	-308	&	235	&	2.6	&	-11	&	11	&	-1.0	&	268	&	11193&11245&$B^{*}_{s}B^{0}_{s1}(5830)$\\
$1^{5}P_{3}$	&	$3^{--}$	&	198	&	-308	&	235	&	2.6	&	21	&	-3.2	&	-1.0	&	249	&	11211&11216&$B^{0}_{s}B^{0}_{sJ}(5850)$\\
$2^{1}P_{1}$	&	$1^{--}$	&	473	&	-218	&	388	&	-4.1	&	0	&	0	&	-0.6	&	306	&	11461&-&-\\
$2^{3}P_{0}$	&	$0^{-+}$	&	473	&	-218	&	388	&	-2.0	&	-17	&	-12	&	-0.6	&	334	&	11433&-&-\\
$2^{3}P_{1}$	&	$1^{-+}$	&	473	&	-218	&	388	&	-2.0	&	-8.5	&	6.3	&	-0.6	&	307	&	11461&-&-\\
$2^{3}P_{2}$	&	$2^{-+}$	&	473	&	-218	&	388	&	-2.0	&	8.0	&	-1.2	&	-0.6	&	297	&	11470&-&-\\
$2^{5}P_{1}$	&	$1^{--}$	&	473	&	-218	&	388	&	2.0	&	-25	&	-14	&	-0.6	&	335	&	11433&-&-\\
$2^{5}P_{2}$	&	$2^{--}$	&	473	&	-218	&	388	&	2.0	&	-8.5	&	-8.0	&	-0.6	&	300	&	11468&-&-\\
$2^{5}P_{3}$	&	$3^{--}$	&	473	&	-218	&	388	&	2.0	&	17	&	-2.5	&	-0.6	&	286	&	11482&-&-\\
[1ex]
\hline
\end{tabular*}
\end{table*}

\begin{table*}[h]
\centering
\caption{The Mass-Spectra of [$bs\bar{b}\bar{q}$], generated from data set IV. (Units are in MeV)}
\label{tab:9}
\footnotesize
\begin{tabular*}{170mm}{@{\extracolsep{\fill}}ccccccccccccc}
\hline
 $N^{2S+1}L_{J}$&$J^{PC}$	&	$\langle E^{0}\rangle$	&	$\langle V^{(0)}_{V}\rangle$	&	$\langle V^{(0)}_{S}\rangle$	&	$\langle V^{(0)}_{SS}\rangle$	&	$\langle V^{(1)}_{LS}\rangle$	&	$\langle V^{(1)}_{T}\rangle$ &	$\langle V^{(1)}(r) \rangle $	&	$\langle K.E. \rangle$	&	$M^{i}_{f}$ & $M_{th}$ &Threshold\\
\hline
$1^{1}S_{0}$	&	$0^{++}$	&	-233	&	-796	&	111	&	-41.0	&	0	&	0	&	-1.9	&	492	&	10522&10645&$B^{\pm}B^{0}_{s}$\\
$1^{3}S_{1}$	&	$1^{+-}$	&	-233	&	-796	&	111	&	-20.5	&	0	&	0	&	-1.9	&	472	&	10573&10690&$B^{*}B^{0}_{s}$\\
$1^{3}S_{1}$	&	$1^{++}$	&	-233	&	-796	&	111	&	0	&	0	&	0	&	-1.9	&	452	&	10594&10690&$B^{*}B^{0}_{s}$\\
$1^{5}S_{2}$	&	$2^{++}$	&	-233	&	-796	&	111	&	20.5	&	0	&	0	&	-1.9	&	431	&	10614&-&-\\
$2^{1}S_{0}$	&	$0^{++}$	&	267	&	-338	&	291	&	-9.3	&	0	&	0	&	-0.9	&	323	&	11085&-&-\\
$2^{3}S_{1}$	&	$1^{+-}$	&	267	&	-338	&	291	&	-4.6	&	0	&	0	&	-0.8	&	319	&	11089&-&-\\
$2^{3}S_{1}$	&	$1^{++}$	&	267	&	-338	&	291	&0	&	0	&	0	&	-0.8	&	314	&	11094&-&-\\
$2^{5}S_{2}$	&	$2^{++}$	&	267	&	-338	&	291	&	4.6	&	0	&	0	&	-0.9	&	309	&	11097&-&-\\
$3^{1}S_{0}$	&	$0^{++}$	&	537	&	-242	&	439	&	-5.8	&	0	&	0	&	-0.6	&	346	&	11358&-&-\\
$3^{3}S_{1}$	&	$1^{+-}$	&	537	&	-242	&	439	&	-3.0	&	0	&	0	&	-0.6	&	343	&	11361&-&-\\
$3^{3}S_{1}$	&	$1^{++}$	&	537	&	-242	&	439	&	0	&	0	&	0	&	-0.6	&	340	&	11364&-&-\\
$3^{5}S_{2}$	&	$2^{++}$	&	537	&	-242	&	439	&	2.8	&	0	&	0	&	-0.6	&	337	&	11367&-&-\\
$1^{1}P_{1}$	&	$1^{--}$	&	202	&	-306	&	237	&	-5.3	&	0	&	0	&	-1.0	&	276	&	11024&11129&$B^{*}B^{*}_{sJ}(5850)$\\
$1^{3}P_{0}$	&	$0^{-+}$	&	202	&	-306	&	237	&	-2.6	&	-22	&	-16	&	-1.0	&	311	&	10988&-&-\\
$1^{3}P_{1}$	&	$1^{-+}$	&	202	&	-306	&	237	&	-2.6	&	-11	&	8.1	&	-1.0	&	276	&	11023&11087&$B_{1}(5721)^{+}B^{0}_{s}$\\
$1^{3}P_{2}$	&	$2^{-+}$	&	202	&	-306	&	237	&	-2.6	&	11	&	-1.6	&	-1.0	&	264	&	11035&11119&$B^{\pm}B^{*}_{s2}(5840)$\\
$1^{5}P_{1}$	&	$1^{--}$	&	202	&	-306	&	237	&	2.6	&	-32	&	-11	&	-1.0	&	312	&	10987&11107&$B^{\pm}B^{*}_{sJ}(5850)$\\
$1^{5}P_{2}$	&	$2^{--}$	&	202	&	-306	&	237	&	2.6	&	-11	&	11	&	-1.0	&	268	&	11032&11136&$B^{*}_{1}(5721)B^{*}_{s}$\\
$1^{5}P_{3}$	&	$3^{--}$	&	202	&	-306	&	237	&	2.6	&	21	&	-3.2	&	-1.0	&	249	&	11050&11155&$B^{*}_{2}(5747)B^{*}_{s}$\\
$2^{1}P_{1}$	&	$1^{--}$	&	478	&	-216	&	391	&	-4.2	&	0	&	0	&	-0.5	&	307	&	11301&-&-\\
$2^{3}P_{0}$	&	$0^{-+}$	&	478	&	-216	&	391	&	-2.1	&	-17	&	-12	&	-0.6	&	335	&	11273&-&-\\
$2^{3}P_{1}$	&	$1^{-+}$	&	478	&	-216	&	391	&	-2.1	&	-8.5	&	6.3	&	-0.6	&	307	&	11301&-&-\\
$2^{3}P_{2}$	&	$2^{-+}$	&	478	&	-216	&	391	&	-2.0	&	8.0	&	-1.2	&	-0.6	&	298	&	11310&-&-\\
$2^{5}P_{1}$	&	$1^{--}$	&	478	&	-216	&	391	&	2.0	&	-26	&	-14	&	-0.6	&	336	&	11272&-&-\\
$2^{5}P_{2}$	&	$2^{--}$	&	478	&	-216	&	391	&	2.0	&	-8.5	&	-8.0	&	-0.6	&	301	&	11307&-&-\\
$2^{5}P_{3}$	&	$3^{--}$	&	478	&	-216	&	391	&	2.0	&	17	&	-2.5	&	-0.6	&	286	&	11322&-&-\\
[1ex]
\hline
\end{tabular*}
\end{table*}

\subsection{Tetraquarks spectrum}
Tetraquarks are color singlet states made up of a diquark $(\mathcal{D})$ and an antidiquark ($\mathcal{\bar{D}}$) in color antitriplet $\mathbf{\bar{3}}$ and triplet $\bf{3}$ configurations respectively, that are held together by color forces \cite{Gyang,debastiani}. The four-body non-relativistic calculation is simplified to a  two-body calculation using this approximation \cite{debastiani,prd,Rohitepja}. A $T_{Qq\bar{Q}\bar{q}}$ is color singlet states and yield a color factor $k_{s} = -\frac{4}{3}$.
The scalar diquark-antidiquark are combined to form color singlet tetraquark \cite{thesis}, and that can be represented as; $|QQ|^{3} \otimes|\bar{Q}\bar{Q}|^{\bar{3}}\rangle = \mathbf{1  \oplus 8}$. The mass-spectra of all bottom ($cq\bar{c}\bar{q}$) and heavy-light strange tetraquarks ($cs\bar{c}\bar{s}$) have been obtained with the same formulation as in the case of mesons, namely;
\begin{equation}
M_{cq\bar{c}\bar{q}} = M_{cq}+ M_{\bar{c}\bar{q}} + E_{[cq][\bar{c}\bar{q}]} + \langle V^{1}(r)\rangle
\end{equation}
and
\begin{equation}
M_{cs\bar{c}\bar{s}} = M_{cs}+ M_{\bar{c}\bar{s}} + E_{[cs][\bar{c}\bar{s}]} + \langle V^{1}(r)\rangle
\end{equation}
The masses of $cs\bar{c}\bar{s}$, $cq\bar{c}\bar{s}$ and $bs\bar{b}\bar{s}$, $bq\bar{b}\bar{s}$ states are anticipated to be in the range of 4.0 - 5.0 GeV, and 10.0 -11.5 GeV respectively in the current study, the masses are also found to be in this range. The obtained masses of these states are also in good accord with two meson thresholds.
The masses of the S-wave $cs\bar{c}\bar{s}$ and $cq\bar{c}\bar{s}$ tetraquark states are slightly higher than the sum of the masses of the two mesons singlets, because strong attraction in color singlet channels is stronger than in color anti-triplet channels. Thus, it is not surprising that the observed tetraquarks appear near to the corresponding meson thresholds, albeit being heavier.
As shown in Tables \ref{tab:6}, \ref{tab:7}, \ref{tab:8}, and \ref{tab:9}, the compactness of the tetraquarks states arise mostly due to the coulomb interaction. This indicates that one-gluon exchange is the dominant mechanism behind the strong interaction between diquarks and antidiquarks, which results in a negative energy eigenvalue E in 1S-wave states. Whereas the contribution of the confinement term increases with the increase in radial and orbital states.
\\
\\
Within the specific tetraquark mass-spectrum, the attractive strength of the spin-spin interaction decreases as the number of radial and orbital excited states increases. In this instance, we must bear in mind that the factors originating from $S_{1}$ and $S_{2}$ are greater for the coupling of two spin-1 particles than for the coupling of two spin-$\frac{1}{2}$ particles. It is worth noting that, despite the fact that the spin-dependent terms have been suppressed by a factor $\frac{1}{m_{qq}^{2}}$, one would anticipate them to be less than the equivalent terms in $q\bar{q}$ mesons.
The color interaction brings diquark and antidiquark so close together that the suppression caused by this component $\frac{1}{m_{qq}^{2}}$, is swamped by the massive suppression at the system's origin. 
The mass of tetraquark states are significantly influenced by orbital excitation. It causes a substantial mass splitting, in the order of 10-100 MeVs, between states with varying orbital angular momentum, owing to the Coulomb interaction and confinement potential. The spin-spin interactions are exceedingly weak in higher orbital excited states, resulting in a mass splitting of few MeV. As a result, the masses of excited states with the same L and S but different total angular momentum J are nearly degenerate, which is qualitatively compatible with the work's conclusion \cite{M. Cleven}. Furthermore, tensor interactions are typically weaker than spin-orbit interactions \cite{M. Furuichi}, which is often overlooked in preliminary research, here in the present model a similar shift have been observed. When the relativistic term $V^{1}(r)$ is incorporated in the central potential, the mass spectrum shifts by a few MeV's. 
\\
\\
The most fundamental multiquark system is the tetraquark system, which consists of two quarks and two antiquarks. The study of heavy tetraquarks is particularly interesting because the presence of a heavy quark increases the binding energy of the bound system, increasing the likelihood that such tetraquarks will have masses below the thresholds for decays to open heavy flavour mesons. The proposed tetraquarks may be considered as having resonances as long as their masses are a few MeV or less above these thresholds. Even though there is a lot of phase space, the excited tetraquark states could be extremely tiny \cite{R. N. Faustov,A. Esposito}. 
\\
\\
In the present model the masses of $cs\bar{c}\bar{s}$ tetraquark matches with several neutral states along with other prediction.
The LHCb Collaboration recently confirmed the states X(4140) and X(4274) in the $J/\psi\phi$ invariant mass distribution, determining their spin-parity quantum numbers to be both $1^{++}$ \cite{LHCbX4140}, which has a significant impact on its conceivable interpretations. The notion of characterising X(4140) as a $0^{++}$ or $2^{++}$ $D^{*+}_{s} D^{*-}_{s}$ molecular state has been ruled out \cite{X. Liu}. At the same time, describing the state X(4274) as a molecular confined state or a cusp is insufficient to account for its quantum numbers \cite{LHCbX4140}.
In Ref. \cite{XinZhen}, authors predicted the mass of $ss\bar{c}\bar{c}$ tetraquark state 4033 MeV, 4192 MeV and 4265 MeV with quantum numbers $0^{+}$, $1^{+}$, and $2^{+}$, respectively and anticipated to be decay in $\bar{D}_{s}\bar{D}_{s}$ and $\bar{D}^{*}_{s}\bar{D}^{*}_{s}$ channel.
\\
\\
The mass of the state $cs\bar{c}\bar{s}$ with $1^{++}$ in this work is 4160 MeV, which is quite near to the state X(4140), roughly 20 MeV above this state. In this approach, the state X(4140) can easily be accommodated as a state $cs\bar{c}\bar{s}$ with $1^{++}$ in the non-relativistic diquark-antidiquark model, but it is nearly 80 MeV above the two meson threshold $D^{\pm}_{s} D^{\pm}_{s}$. However, the mass of X(4274) can be compared with the mass of the 1S-state with $2^{++}$ though its is 50 MeV below from X(4274), implying that the state  major component of X(4274) could be the state $cs\bar{c}\bar{s}$ with $2^{++}$ in $D^{*\pm}_{s} D^{*\pm}_{s}$. Many prior studies of these two states have found that arranging the two states in the same theoretical framework under the assumption of $1^{++}$ is difficult \cite{R. F. Lebed,M. Padmanath}. QCD sum rules and simple color-magnetic interaction models, on the other hand, can both interpret X(4140) and X(4274) as S-wave states with $1^{++}$ \cite{Fl Stancu,J.Wu,W. Chen2011}.
\\
\\
In addition to the states X(4140) and X(4274), the high $J/\psi\phi$ mass region has been studied for the first time with high sensitivity and exhibits highly significant structures, the states X(4500) and X(4700), which can be regarded as two $0^{++}$ resonances \cite{LHCbX4140}. 
When comparing the data, the mass of the lowest S-wave [cs][$\bar{c}\bar{s}$] state with $0^{++}$ in the current work appears to be too light, but the radial excitation 2S-wave, which satisfies the quantum number requirement and has a mass of nearly X(4700) but is appears 200 MeV far from the X(4500). The idea of these two states as a P-wave or D-wave has been ruled out in Ref. \cite{C. Deng} but expected to have fourth and fifth S-wave radial excited state. The two states were also explained by Zhu \cite{R. L. Zhu} as radial excitation of the $J^{P}=0^{+}$ tetraquark state. Within the framework of QCD sum rules, Chen et al. \cite{H. X. Chen2017} regarded the two states as the D-wave [cs][$\bar{c}\bar{s}$] tetraquark states of $J^{P}=0^{+}$. In addition to the [cs][$\bar{c}\bar{s}$] explanation, the states X(4500) and X(4700) were described as standard charmonium states with $4^{3}P_{1}$ and $5^{3}P_{1}$ that match the states X(4500) and X(4700) respectively, in the non-relativistic constituent quark model \cite{P. G. Ortega}.
In two-photon collisions, the Belle Collaboration reported a small $J/\psi\phi$ peak at 4350.6 $ ^{4.6} _{5.1} \pm0.7 $ MeV, implying $J^{PC}=0^{++}$ or $2^{++}$ \cite{C. P. Shen}. In Table VI shows that the pure $[cs]_{\bar{3}_{c}}[\bar{c}\bar{s}]_{3_{c}}$ state with $J^{PC}=0^{++}$ is inconsistent with X(4350),whereas $2^{++}$ has a mass of 4218 MeV, which is in line with the experimental findings. If $J^{PC}=2^{++}$ is correct, our forecast of 4218 MeV is slightly lower than the experiment's result but closer to $D^{*\pm}_{s}D^{*\pm}_{s}$ meson threshold.
\\
\\
In several processes, Belle Collaboration observed the states Y(4626), Y(4630), and Y(4660) with $1^{--}$ \cite{S. Jia,X. L.,G. Pakhlova}. Their masses and widths, on the other hand, are within errors of each other. To illustrate the internal structure of the Y(4660) and Y(4630), many theoretical explanations have been done, in which the two states were understood as the same state \cite{D. V. Bugg} the tetraquark state [$cq][\bar{c}\bar{q}$] and $f_{0}(980)\psi^{'}$ bound state.

The mass of the P-wave tetraquark state $[cs]^{1}$ [$\bar{c}\bar{s}]^{1}$ with $1^{--}$ for $S_{T}=0$, in the current work is 4556 MeV, which is consistent with that of the state Y(4660). The masses of Y(4660) along with other states namely Y(4626), and Y(4630) are consistent with P-wave tetraquarks with quantum number $1^{--}$ and can be characterised as a two meson threshold mentioned in Table 6. 
Since both Y(4630) and Y(4626) have the same spin structure, which consists of a scalar [cs] and a scalar [$\bar{c}\bar{s}$] \cite{C. Deng}, on the other hand they are considered as a vector state in this work. On the other hand, the result of the QCD sum rules \cite{R. M. Albuquerque} suggests that the state Y(4660) is composed of a scalar [cs] ($\bar{c}\bar{s}]$) and an axial-vector [$\bar{c} \bar{s}$] ([cs]). According to Ref. \cite{C. Deng}, the P-wave state $[[cs]_{\bar{3}_{c}}[\bar{c}\bar{s}]_{3_{c}}]_{1}$  with $1^{--}$ has a mass of 4704 MeV, which is close to the mass of the Y(4660) state. The colour configuration $[[cs]_{\bar{3}_{c}}[\bar{c}\bar{s}]_{3_{c}}]_{1}$ in all three states is a significant advantage; alternatively, the colour configuration $[[cs]_{\bar{6}_{c}}[\bar{c}\bar{s}]_{6_{c}}]_{1}$ can be disregarded. In reality, the model predicts that the three states [$cs][\bar{c}\bar{s}$] with $1^{--}$ can interact through the tensor interaction, which will be calculated with greater precision in the future. The predictions from our work are very close with Ref. \cite{R. N. Faustov}, where Faustov et al., predicted the masses of $[cs][\bar{c}\bar{s}]$, $[cq][\bar{c}\bar{s}]$ and $[bs][\bar{b}\bar{s}]$, $[bq][\bar{s}\bar{s}]$ tetraquark states.
The $Z_{cs}$(3985) state \cite{M Ablikim et al}, which has been observed as $1^{+}$ state \cite{R. N. Faustov}, can be accommodated as $1^{+-}$ state in our model (tabulated in \ref{tab:7}.) as $D^{\pm}D^{*\pm}_{s}$ meson threshold . Along with that other states of [$cq\bar{c}\bar{s}$] matches with several two-meson thresholds which is tabulated in Table \ref{tab:7}, could be a subject of future experimental prediction.
\\
\\
In the multiquark colour flux-tube model \cite{C. R. Deng}, the masses of the P-wave states [$bs][\bar{b}\bar{s}$] with $1^{--}$ are calculated to be between 11,099 MeV and 11,134 MeV. In the present work, masses of [$bs][\bar{b}\bar{s}$] tetraquark states are found to be in the range 10.700 GeV- 11.500 GeV. 
In Ref. \cite{mb} mass of ground state [$bs][\bar{b}\bar{s}$] has been obtained 10.52(08) GeV in scalar-scalar diquark-antiquark configuration. In Ref. \cite{XinZhen}, author predicted the ground state masses of [$ss][\bar{b}\bar{b}$] tetraquark state which are expected as stable tetraquarks. The masses of two quantum states $0^{+}$ and $1^{+}$, are 10697 MeV and 10718 MeV respectively which lies below the $B_{s}B_{s}$ threshold whereas, for $2^{+}$ state its around 10742 MeV, this lies just above the $B_{s}B_{s}$ threshold.

Although no bound state has been discovered in the two types of tetraquarks: [$cc][\bar{s}\bar{s}$] and [$bb][\bar{s}\bar{s}$], various resonances have been predicted \cite{Gyang}. In the present model, with $J^{PC}$ = $0^{++}$ quantum numbers, a narrow di-$D^{+}_s$ resonance is identified for the [$cs][\bar{c}\bar{s}$] tetraquark, with expected mass of 4045 MeV. In addition, several resonance having $D^{+}_sD^{+}_s$ meson threshold with masses of roughly 4.1-4.6 GeV have been predicted and listed in \ref{tab:6}. The [$bs][\bar{b}\bar{s}$] tetraquark sector, whose resonances are exclusively found in $0^{++}$ and $1^{++}$, yields similar results. In the P-wave state, there are four channels which are compared with two meson thresholds. Their masses are in the 11.1 – 11.2 GeV range. 
In ref. \cite{R. N. Faustov}, author have computed the  masses of tetraquark state [$bq\bar{b}\bar{s}$] between 10.5 GeV- 11.3 GeV, which is fairly close to our predictions.
\section{Summary}
We investigate the tetraquark states with the diquark-antidiquark picture in the multiquark non-relativistic model with a Cornell potential and other spin-dependent interactions systematically. There are one gluon confinement potential along with linear term that drives the formation of bound or resonance states. A significant mixing of diquark and antidiquark colour configurations occurs in the ground state, while the colour configuration dominates in the excited states and is completely dominant. The most discussed tetraquark state  X(4140), which can be explained as a S-wave state [$cs\bar{c}\bar{s}$] with quantum number $1^{++}$ and mass difference of 50 MeV.
The P-wave tetraquark states [$cs\bar{c}\bar{s}$] with $J^{PC}$=$1^{--}$ can accommodate Y(4626), Y(4630), and Y(4660) tetraquark states whereas X(4274) as ground state tetraquark $2^{++}$ with two-meson threshold $D^{*\pm}_{s} D^{*\pm}_{s}$.
\\
\\
In theory, one gluon-exchange (OGE) force may bind four quark states. However, since there is a dearth of trustworthy and unambiguous experimental evidence, their likely development and stability is debatable. As a result, tetraquark model predictions are very model dependent and reliant on the Hamiltonian used as well as the parameter fitting technique. Regardless, the tetra-quark idea is worth exploring.

%



\end{document}